# Effect of Thermal Phase Fluctuations on the Inductances of Josephson Junctions, Arrays of Junctions, and Superconducting Films

Thomas R. Lemberger, Aaron A. Pesetski, and Stefan J. Turneaure
*Dept. of Physics*
*Ohio State University*
*Columbus, OH 43210-1106*

We calculate the factor by which thermal phase fluctuations, as distinct from phase-slip fluctuations, increase the inductance, $L_J$, of a resistively-shunted Josephson junction (JJ) above its mean-field value, $L_0$. We find that quantum mechanics suppresses fluctuations when T drops below a temperature, $T_Q = \hbar/k_B G L_0$, where G is the shunt conductance. Examination of the calculated sheet inductance, $L_A(T)/L_0(T)$, of arrays of JJ's reveals that 2-D interconnections halve fluctuation effects, while reducing phase-slip effects by a much larger factor. Guided by these results, we calculate the sheet inductance, $L_F(T)/L_0(T)$, of 2-D films by treating each plasma oscillation mode as an overdamped JJ. In disordered s-wave superconductors, quantum suppression is important for $L_F(0)/L_F(T) > 0.14$, (or, $T/T_{C0} < 0.94$). In optimally doped YBCO and BSCCO quantum suppression is important for $\lambda^2(0)/\lambda^2(T) > 0.25$, where $\lambda$ is the penetration depth.

1. Introduction.

This work is directed toward understanding the effect of thermal phase fluctuations on the sheet inductance of two-dimensional (2-D) homogeneous superconducting films. Strictly speaking, thermal phase fluctuations include phase-slip fluctuations, *i.e.*, vortex-antivortex pairs, which are generated near the 2-D transition where fluctuations are very strong, and whose dynamics are believed to mediate the super-to-normal transition as measured resistively. In this work we are interested in the non-phase-slip part of the fluctuations, and that is what we mean by, "thermal phase fluctuations." (These are sometimes called phase phonons or spin waves.) A key issue naturally involves the width of the vortex-dominated region: Over what range of temperatures does the effect of vortex-antivortex pairs on the sheet inductance exceed that of thermal phase fluctuations? Another question involves lower temperatures: Below what temperature, and by what factor, does quantum mechanics suppress thermal phase fluctuations?

The superconducting-to-normal phase transition in two dimensions is a problem of longstanding interest [1,2]. In films of conventional s-wave superconductors, and in arrays of Josephson junctions, the dynamics of vortex-antivortex pairs as calculated in the Kosterlitz-Thouless-Berezinski (KTB) theory does not describe the dependencies of the complex sheet resistivity, $\rho/d = \rho_1(\omega,T)/d - j\rho_2(\omega,T)/d$, on $\omega$ or T particularly well [3-5]. In particular, the KTB theory predicts that the contribution of vortex-antivortex pairs to the sheet inductance should be confined to a very narrow temperature interval, just below the transition, in which the spacing between pairs and the size of a typical pair are comparable, while the experimentally observed interval over which fluctuation effects are evident is much larger. The present work finds that, as regards $\rho_2$, vortex-antivortex pairs are indeed important only very close to the transition, and thermal phase fluctuations account for most of the upturn in $\rho_2$, *i.e.*, the downturn in areal superfluid density, $n_S(T) \propto d/\lambda^2(T) \equiv 1/\lambda_\perp(T)$, that occurs as T approaches $T_{KTB}$ [3-6].

A more recent problem concerns the role of thermal phase fluctuations in quasi-2-D cuprate superconductors. A quick analysis shows that a KTB transition would occur in each CuO layer of optimally-doped YBCO about 5 K below the measured $T_C$, if there were no coupling between layers. This suggests that fluctuations could be the dominant influence on the T-dependence of the superfluid density over a much wider temperature range than we are familiar with from studies of thin films of low-$T_C$ superconductors. Indeed, the penetration depth, $\lambda^{-2}(T)$, [7,8] and thermal expansivity [9] measured in very clean YBCO crystals seem to exhibit critical fluctuations over a 5 to 10 K interval up to $\approx 0.998\ T_C$. The situation is a bit clouded for several reasons. First, the critical exponent indicates that fluctuations are 3-D, not 2-D. Second, the width of the critical region is very sensitive to unknown parameters: $\lambda^{-2}(T)$ measured[10] on YBCO crystals which are nominally identical to those of refs. [7,8], and on high-quality YBCO films [11,12], do not exhibit critical fluctuations. Despite these concerns, the possible significance of phase fluctuations in cuprates must be explored. Qualitatively, phase fluctuations can account for [13-16] the T-linear behavior of $\lambda^{-2}(T)$ at low T [17,18], and for the approximate proportionality between $T_C$ and $\lambda^{-2}(0)$ for underdoped cuprates[19].

In the present work, we lay the groundwork for a critical examination of phase fluctuations in cuprates by considering their role in simpler systems, namely, Josephson junctions and arrays of junctions. We apply our results to the question of whether thermal phase fluctuations could account for the T-linear behavior in λ in cuprates at low T, and defer the more complicated question of critical behavior near $T_C$.

The most detailed calculation of the effect of phase fluctuations on the sheet inductance of a superconducting film is that of Coffey [16], who calculated the lowest-order effect of classical phase fluctuations within the Lawrence-Doniach model [20]. The present work can be viewed as an extrapolation of that work to higher temperatures, where fluctuations are large and nonlinear effects come into play, and lower temperatures where quantum mechanics is important. Our calculation employs some approximations, but our final result for films is consistent with measurements of the sheet inductance of thin homogeneous films of a conventional superconductor, amorphous MoGe, including the quantum crossover [5]. An interesting unconventional measurement of thermal phase fluctuations is the tunneling study of weakly disordered, thin, homogeneous superconducting Al films [21]. The decrease in the relaxation time of a quasiparticle charge imbalance with increasing sheet resistance of Al films agreed well with the present model, although possible quantum effects were not considered.

2. Inductance of a Josephson Junction.

Quantum suppression of thermal phase fluctuations is of central importance, but it is difficult to calculate for a homogeneous film. Fortunately, it emerges naturally from a calculation of the effect of thermal phase fluctuations on the inductance, $L_J(T)$, of a Josephson tunnel junction (JJ).

The physics of JJ's is described in detail by Likharev [22]. The simplest model has a junction with intrinsic critical current, $I_C(T)$, in parallel with a capacitance, C, (Fig. 1), and an external shunt resistor, R, that is much smaller than the normal-state resistance of the junction. For low amplitude *ac* bias current, and in the absence of noise, the junction behaves like an inductor with impedance $j\omega L_0 = j\omega\hbar/2eI_C(T)$. The shunt ensures that the effective junction resistance is independent of the voltage across it, thereby simplifying the equation of motion for the phase difference across the junction. It also ensures that junction dynamics take place below a low-pass frequency, $\omega_0(T) \equiv R/L_0(T)$, that is much smaller than the gap frequency, $\Delta(T)/\hbar$, in the junction electrodes. With this constraint, $I_C$ is effectively independent of ω. C is the sum of the physical capacitance of the junction and the effective quasiparticle capacitance obtained from the Kramers-Kronig (KK) transform of the quasiparticle contribution to the real conductance, $\sigma_1(\omega)$, of the tunnel junction [23].

Thermal fluctuations originate in the resistor and are represented by a noise current, $i_n(t)$, in parallel with R. The influence of thermal noise on the junction comes from the mean square supercurrent, $\langle I_S^2 \rangle$, through the junction. We neglect fluctuations in $I_C$. It is straightforward to see why only the low-frequency components of $i_n$ contribute to $\langle I_S^2 \rangle$. Noise currents are "white" up to $\omega \approx k_BT/\hbar$ and diminish at higher ω due to quantum mechanics [24]. Low frequency noise currents pass through the junction as supercurrents because the impedance of the junction is much less than the resistor or capacitor, *i.e.*, $\omega L_0 \ll \{R, 1/\omega C\}$. As $\omega$ increases, eventually $\omega L_0$ exceeds either R or $1/\omega C$, and noise currents pass through the resistor or capacitor instead of the junction. For an overdamped junction, R is much less than $1/\omega C$, and only noise currents with $\omega < \omega_0$ contribute to $\langle I_S^2 \rangle$. Thus, when $k_B T/\hbar$ drops below $\omega_0$, $\langle I_S^2 \rangle$ drops below its classical value. This is what is meant by quantum suppression of thermal fluctuations.

Calculation of the normalized inductance, $L_J(T)/L_0(T)$, of the junction proceeds as follows. [22] With an external bias current, $I_b(t) = I_0 + I_{ac}(t)$, which includes a small *ac* component, $I_{ac}$, at angular frequency $\omega$, conservation of current leads to:

$$I_0 + I_{ac}(t) + i_n(t) = I_C \sin[\phi(t)] + GV(t) + C\, dV(t)/dt. \qquad 1$$

In Eq. (1), $G \equiv 1/R$ is the conductance of the resistor. With the Josephson relation, $d\phi(t)/dt = 2eV(t)/\hbar$, for the phase difference, $\phi(t)$, across the junction, a time derivative leads to:

$$dI_{ac}(t)/dt + di_n(t)/dt = I_C \cos(\phi) 2eV(t)/\hbar + G\, dV(t)/dt + C\, d^2V(t)/dt^2 \qquad 2$$

Taking an ensemble average, we find to lowest order in thermal noise:

$$dI_{ac}(t)/dt \approx I_C \langle\cos(\phi)\rangle 2eV(t)/\hbar + G\, dV(t)/dt + C\, d^2V(t)/dt^2 \qquad 3$$

$$\approx I_C [1 - (I_0^2 + \langle I_S^2 \rangle)/I_C^2]^{1/2}\, 2eV(t)/\hbar + G\, dV(t)/dt + C\, d^2V/dt^2. \qquad 4$$

Small *ac* bias means: $I_{ac}/\omega \ll I_C/\omega_0 = 2eG/G_Q$, ($G_Q \equiv 4e^2/\hbar \approx 1/1027\ \Omega$). $\langle I_S^2 \rangle$ is the mean square supercurrent through the junction. A Fourier transform yields:

$$I_{ac}(\omega) = V(\omega)[1/j\omega L_J + G + j\omega C]. \qquad 5$$

Thus, $L_J$ increases with *dc* bias, $I_0^2/I_C^2$, and with thermal noise, $\langle I_S^2 \rangle/I_C^2$:

$$L_0(T)/L_J(I_0,T) = [1 - I_0^2/I_C^2 - a\langle I_S^2 \rangle/I_C^2]^{1/2}. \qquad 6$$

"*a*" measures the sensitivity of the inductance to thermal supercurrents. The preceding analysis leads us to expect $a \approx 1$ for Josephson junctions. We work in terms of the inverse inductance, as in Eq. (6), because $1/L_J$ is analogous to the superfluid density in a film. When $\langle I_S^2 \rangle$ is sufficiently large, the phase difference across the junction can slip by $2\pi$ resulting in a small voltage spike. In order to understand the importance of phase-slip events relative to thermal phase fluctuations, we calculate the full junction impedance below.

Before calculating $L_J(T)$ for an unbiased junction from Eq. (1), we pause to calculate $\langle I_S^2 \rangle$ *vs.* T and quantify the quantum crossover. The mean square noise current that lies within a small bandwidth, $\Delta B$, centered on $\omega$ is [24]:

$$\langle |i_n(\omega)|^2 \rangle \Delta B = 4G \{\hbar\omega/2 + \hbar\omega/[e^{\hbar\omega/kT} - 1]\} \Delta B \qquad 7$$

The *rhs* of Eq. (7) reduces to the classical value, $4k_BTG\Delta B$, for $\hbar\omega \ll k_BT$. From here on, we neglect the zero-point motion quantum contribution, $2G\hbar\omega\Delta B$, to $\langle |i_n(\omega)|^2 \rangle$. The mean square *thermal* noise current through the junction is

$$\langle I_S^2 \rangle = \int_0^\infty (d\omega/2\pi)\, 4G\, \hbar\omega[e^{\hbar\omega/k_BT} - 1]^{-1} [\omega^2/\omega_0^2 + (1 - \omega^2/\omega_J^2)^2]^{-1} \qquad 8$$

where $\omega_J \equiv (L_0 C)^{-1/2}$. In the classical limit, $k_BT/\hbar \gg \min\{\omega_0, \omega_J\}$, Eq.(8) yields the classical result: $\langle I_S^2 \rangle = k_BT/L_0 \equiv \langle I_S^2 \rangle_C$, by direct integration. The thermal factor in Eq. (8) shows that when $k_BT/\hbar$ drops below $\min\{\omega_0, \omega_J\}$, only a portion, $\approx k_BT/\hbar\, \min\{\omega_0, \omega_J\}$, of the low-pass band is excited, and we expect $\langle I_S^2 \rangle$ to drop below its classical value by about this factor.

We define the noise parameter as: $\gamma(T) \equiv \langle I_S^2 \rangle / I_C^2$, and calculate it from Eq. (8). In the classical limit, $\gamma(T) = k_BT/J(T)$, where the characteristic energy, $J(T) \equiv \hbar/G_Q L_0(T)$, is the usual Josephson coupling energy. If we define: $\gamma_0 \equiv k_BT/J(T)$, then we can write $\gamma = \gamma_0 f_Q$, where the quantum suppression factor is: $f_Q \equiv \langle I_S^2 \rangle / \langle I_S^2 \rangle_C$. Figure 2 shows $f_Q$ *vs.* $k_BT/\hbar\omega_0$, calculated for $\omega_J \gg \omega_0$, $\omega_J = 0.25\omega_0$, and $\omega_J = 0.70\omega_0$. The dotted curves in Fig. 2 show approximations:

$$f_Q \approx 1/(1 + \hbar\omega_0/k_BT) \qquad 9a$$

for the overdamped case, $\omega_J \gg \omega_0$, and

$$f_Q \approx (2/\pi)\arctan(\pi k_BT/2\hbar\omega_J), \qquad 9b$$

for the underdamped cases, $\omega_J = 0.25\,\omega_0$ and $0.7\,\omega_0$. We define a quantum crossover temperature, $T_Q$, from the condition:

$$k_BT_Q = \hbar\, \min\{\omega_0(T_Q), \omega_J(T_Q)\}. \qquad 10$$

For $T/T_Q < 1/2$,

$$\gamma \approx \gamma_0\, k_BT/\,[\hbar\, \min\{\omega_0(T), \omega_J(T)\}], \qquad 11$$

for overdamped and underdamped junctions. The formal quadratic T dependence of $\gamma$ ($\omega_0$ and J depend on T, too) at low T was pointed out by Millis et al. [25].

We emphasize that for overdamped junctions the "R/L" frequency, $\omega_0$, is important and the plasma frequency, $\omega_J$, is not. We argue below that the same is true for homogeneous superconducting films because the quantum crossover occurs at high temperatures where plasma oscillations are highly damped. In principle, the plasma frequency can be important at low temperatures, where $\sigma_1(\mathbf{q},\omega,T)$ is very small and damping is weak.

Following Likharev [22], the impedance of an overdamped junction is obtained from the solution to the Smoluchowski equation for the probability, $\sigma(\phi,t)$, for the junction to sustain at time $t$ a phase difference $\phi$, when the normalized bias current is: $I(t) = I_0 + I_{ac}\sin(\omega t)$:

$$(1/\omega_0)\partial\sigma(\phi,t)/\partial t + \partial\{\sigma(\phi,t)[\,I_0 + I_{ac}\sin(\omega t) - \sin(\phi)]\}/\partial\phi = \gamma_0\partial^2\sigma/\partial t^2. \qquad 12$$

To solve, $\sigma(\phi,t)$ is Fourier transformed in both $\phi$ and t, with the assumption that $\sigma(\phi,t)$ is periodic in $\phi$ with period $2\pi$. The Fourier components of $\sigma$ that involve $e^{\pm j\omega t}$ are calculated, and from them the junction impedance, $Z_J = R_J + j\omega L_J$, is deduced.

Figure 3 shows $L_J/L_0$ *vs.* $\gamma_0$ and $R_J/R$ *vs.* $\gamma_0$ for an unbiased junction ($I_0 = 0$) at several frequencies, $\omega/\omega_0$. In the present work, we are interested in low frequencies, $\omega/\omega_0 \ll 1$, at which the resistance, $R_J$, necessarily arises from phase slip events. In this limit, $R_J/R$ is found to be tiny for $\gamma_0 < 0.20$, indicating that phase-slip events are rare. Therefore, the increase in $L_J/L_0$ for $\gamma_0 < 0.20$ comes from thermal phase fluctuations, not phase slips. For later comparison with arrays and continuous films, Fig. 4 shows $L_0/L_J$ *vs.* $\gamma_0$ for an overdamped junction in the classical limit. The dotted line shows that the function, $[1 - \gamma_0]^{1/2}$, fits the numerical results for small $\gamma_0$, *i.e.*, $\gamma_0 < 0.20$, as anticipated in Eq. (6).

3. Arrays of Identical Josephson Junctions.

We seek to understand the effect of thermal (non-phase-slip) phase fluctuations on the inverse sheet inductance, $1/L_A$, of a square JJ array. In a film, this quantity is proportional to the superfluid density. In the classical limit, the current noise, $\gamma(T) \equiv \langle I_S^2\rangle/I_C^2$, in each junction is set by the Equipartition Theorem, and is unaffected by interconnections. We expect, and find, that the 2-D array is affected by phase-slip fluctuations much less than a 0-D single junction, permitting a super-to-resistive phase transition, instead of a crossover.

The sheet inductance, $L_A(T)$, of an array of identical, *noninteracting* inductors, L, is proportional to L. (For a square array, $L_A = L$. For triangular and honeycomb arrays, $L_A = L/\sqrt{3}$ and $\sqrt{3}L$, respectively.) The proportionality is lost for an array of identical Josephson junctions because of noise. In an array, the noise current in each junction comes from all of the shunt resistors, not just the local shunt. The inductances of nearby junctions fluctuate in a correlated way, so that they are effectively interacting inductors.

Figure 4 shows the normalized inverse sheet inductance, $L_0/L_A$ *vs.* $\gamma_0$, calculated for a square array of overdamped JJ's in the classical limit [26,27]. In arrays, we define: $\gamma_0 \equiv k_B T\, G_Q\, L_0(T)/\hbar$, where $L_0$ is the mean-field sheet inductance of the array. As seen in Fig. 4, the function, $(1 - a\gamma_0)^{1/2}$ with $a = \frac{1}{2}$ (dotted curve), fits the numerical data for $\gamma_0 <$ 0.7. If $L_A$ were proportional to $L_J$, then $a$ would be unity. We interpret the agreement between this simple function and the exact calculation to mean that for $\gamma_0$ up to at least 0.7, thermal phase fluctuations dominate over phase-slip fluctuations, and the 2-D interconnections among junctions halve their effect. Calculations for triangular and honeycomb arrays[28] lead to similar values, namely, $a = 1/1.7$ and $1/2.5$, respectively. Thus, the influence of thermal phase fluctuations on 2-D arrays is relatively insensitive to details. The intersection of the line labeled "$2\gamma_0/\pi$" with the curve for $L_0/L_A$ *vs.* $\gamma_0$ marks the KTB transition where, in principle, $L_0/L_A$ drops discontinuously from $2\gamma_0/\pi$ to zero.

Phase slip fluctuations in the form of vortex-antivortex pairs account for some of the difference between the numerical data and the approximation, $(1 - a\gamma_0)^{1/2}$, which represents thermal phase fluctuations. Even if all of the difference were due to vortex-antivortex pairs, the suppression of the "superfluid density", $L_0/L_A$, would still be dominated by thermal phase fluctuations except for temperatures very close to the transition.

4. Homogeneous Superconducting Films.

We now consider homogeneous 2-D films. We calculate $L_0/L_F$, where $L_0(T)$ is the "mean-field" sheet inductance. $1/L_F(T)$ is proportional to the areal superfluid density, $n_S(T)$. The calculation is approximate, but it provides insight into how microscopic details would enter a more rigorous calculation. The most serious approximation, in our view, is the omission of fluctuations in the amplitude of the order parameter, which should become significant near the 2-D super-to-resistive transition.

Equation (13) is the analog of Eq. (6) for a Josephson junction and it describes how the sheet inductance is affected by supercurrent fluctuations:

$$L_0(T)/L_F(T) \approx 1 - a(T)\langle p_s^2\rangle \xi(T)^2/\hbar^2 \equiv 1 - \gamma. \qquad 13$$

While Eq. (13) is most easily derived within G-L theory [16,29], it is generally valid. $\langle p_s^2\rangle$ is the mean square thermal momentum of a Cooper pair, and $\langle p_s^2\rangle \xi^2/\hbar^2 \approx \langle \phi^2\rangle$, the mean square spatial variation in phase of the order parameter.

The factor $a(T)$ is a measure of how strongly phase fluctuations affect the sheet inductance. For dirty limit superconductors, $a(T)$ is temperature independent and of order unity. We expect $a$ to be less than unity in analogy with the noise term found for arrays where the suppression of the sheet inductance was proportional to $(1 - \gamma_0/2)^{1/2}$ ($a = 1/2$) in contrast to single junctions where $a = 1$. For clean superconductors, $a(T)$ should be unity near $T_C$, but have a strong T dependence at low T. In the end, $a$ for dirty limit superconductors must be decided by experiment.

Calculating $\gamma$ amounts to calculating $\langle p_s^2 \rangle$, which we do by summing $\langle p_s^2 \rangle_\mathbf{k}$ over plasma oscillation modes, labeled by a wavevector $\mathbf{k}$, and by treating each mode like an overdamped JJ, in analogy with Eq. (8). We cut off the sum on $\mathbf{k}$ for $|\mathbf{k}| > 2\pi/\xi(T)$, presuming that the superfluid is insensitive to fluctuations at length scales shorter than $\xi(T)$. To use Eq. (8), we must connect the fluctuation current of each mode, $\langle I_S^2 \rangle_\mathbf{k}$, with $\langle p_s^2 \rangle_\mathbf{k}$. The inductance and conductance in Eq. (8) connect with the sheet conductance, $\sigma d = \sigma_1 d - j\sigma_2 d$, of a film. We define $\sigma_{2S}(\mathbf{k},\omega,T)$ to be the Kramers-Kronig transform of the delta function in $\sigma_1(\mathbf{k},\omega,T)$ at $\omega = 0$, so $\sigma_{2S}(\mathbf{k},\omega,T) \propto 1/\omega$, and we can define a sheet inductance $L_F(\mathbf{k},T)$ as: $L_F(\mathbf{k},T) \equiv 1/\omega\sigma_{2S}(\mathbf{k},\omega,T)d$. With generic film dimensions, $W \times W \times d$, we have:

$$\langle I_S^2 \rangle_\mathbf{k} = W^2 d^2 \langle J_S^2 \rangle_\mathbf{k} = (n_S e W d/2m)^2 \langle p_s^2 \rangle_\mathbf{k} = W^2 \langle p_s^2 \rangle_\mathbf{k} / [L_F(\mathbf{k},T) 2e]^2 \qquad 14$$

$J_S = n_S e p_S/2m$ is the supercurrent density, and $2m$ is the mass of a Cooper pair. The shunt conductance, $G$, in Eq. (8) becomes $\sigma_1(\mathbf{k},\omega,T)d$. We neglect capacitance by setting $\omega_J$ to infinity. We assume that the thermal factor which represents noise currents is the same as for a JJ. Replacing the lumped circuit elements implicit in $\omega_0$ in Eq. (8) by corresponding parameters for the film, the "circuit" factor becomes:

$$[\omega^2/\omega_0^2 + 1]^{-1} \to [(\omega\sigma_1 L_F d)^2 + 1]^{-1}. \qquad 15$$

We find:

$$\gamma \approx a(T) {\sum}'_\mathbf{k} [\xi L_F(\mathbf{k},T) 2e/\hbar W]^2 \int_0^\infty d\omega [2\hbar\omega\sigma_1 d/\pi][e^{\hbar\omega/kT} - 1]^{-1} [(\omega\sigma_1 d L_F)^2 + 1]^{-1}. \qquad 16$$

The prime on the summation indicates a cutoff at $|\mathbf{k}| = 2\pi/\xi(T)$.

We can approximate the sum in Eq. (16) because $\mathbf{k}$-dependence is generally unimportant. For disordered s-wave superconductors, $\sigma$ is independent of $\mathbf{k}$. For d-wave superconductors, the dependence of $\sigma$ on $\mathbf{k}$ is not well known, but most terms in the sum over $\mathbf{k}$ have $|\mathbf{k}| \approx \sqrt{2}\pi/\xi$, and for the nearly tetragonal *ab*-plane of cuprates we expect $\sigma$ to be more sensitive to the magnitude than to the direction of $\mathbf{k}$. That is, in Eq. (16), $\sigma \approx \sigma(|\mathbf{k}| \approx \sqrt{2}\pi/\xi, \omega, T)$. To evaluate Eq. (16), we replace the sum on $\mathbf{k}$ by the number of terms in the sum, $(W/\xi)^2$, times a single "average" term in which parameters represent the appropriate averages. We replace $\sigma_1(|\mathbf{k}| \approx \sqrt{2}\pi/\xi, \omega, T)d$ by $G_F(T)$, where $\sigma_1$ is averaged over frequencies up to $2\Delta(T)/\hbar$, and we replace $L_F(|\mathbf{k}| \approx \sqrt{2}\pi/\xi, T)$ by $L_F(T)$. With these approximations the noise term may be written as:

$$\gamma \approx a(T) [k_B T\, G_Q\, L_F(T)/\hbar]\, f_Q(T)$$

$$= a(T)\, \gamma_0\, (L_F/L_0)\, f_Q(T), \qquad 17$$

where Eq. (9a) may be used for $f_Q(T)$. $\gamma_0 \equiv k_B T\, G_Q\, L_0/\hbar$ is the classical value of $\gamma$.

We can identify the normalized superfluid density at the quantum crossover from the equation:

$$k_B T_Q = \hbar / G_F(T_Q) L_F(T_Q). \qquad 18$$

Anticipating that $T_Q$ is close to $T_C$, we set $T_Q = T_C$ on the *lhs*. From the conductivity sum rule, [29] $G_F(T)$ is approximately equal to its value, $1/R_N$, just above $T_C$, multiplied by the normal-fluid fraction, $1 - n_S(T)/n_S(0) = 1 - L_F(0)/L_F(T)$:

$$R_N G_F(T) \approx 1 - L_F(0)/L_F(T). \qquad 19$$

If we define a characteristic "R/L" temperature, $T_0 \equiv \hbar R_N / k_B L_F(0)$, then Eqs. (18) and (19) predict a crossover at:

$$n_S(T_Q)/n_S(0) = L_F(0)/L_F(T_Q) \approx (1 + T_0/T_C)^{-1}. \qquad 20$$

As discussed in the following sections, for cuprates and for dirty s-wave superconductors, $T_0$ is several times larger than $T_C$, so $n_S$ is much smaller than $n_S(0)$ at the crossover. To estimate $T_Q$ from Eq. (20), we use the approximation: $L_F(0)/L_F(T_Q) \approx 3(1 - T_Q/T_C)$, which is valid near $T_C$, to obtain:

$$1 - T_Q/T_C \approx 0.33/(1 + T_0/T_C). \qquad 21$$

Below $T_Q$ the noise term is then:

$$\gamma \approx a(T) \, (k_B T)^2 \, G_Q G_F(T) L_F^2(T)/\hbar^2. \qquad (T/T_Q < \tfrac{1}{2}) \qquad 22$$

To compare our result for films with previous results on arrays, we examine the classical limit ($f_Q = 1$). Since our calculation does not improve on the order of magnitude uncertainty in $\gamma$ in the literature, we choose $a(T) = \tfrac{1}{4}$ in Eq. (17) so that for small fluctuations, $\gamma_0 \ll 1$, $L_0/L_F$ agrees with $L_0/L_A$ calculated for square arrays. Coffey calculates a slightly smaller value: $a(T) = \tfrac{1}{4} \ln(2)/\pi$.[16] With these assumptions, Eqs. (13) and (17) yield:

$$L_0/L_F \approx 1 - \gamma_0 L_F/4 L_0. \qquad \text{(Classical limit)} \qquad 23$$

In our calculation, $\gamma$ depends on the film's fluctuation-enhanced sheet inductance, so that Eq. (23) includes nonlinear effects from strong fluctuations. Solving Eq. (23) for $L_0/L_F$ as a function of the normalized temperature, $\gamma_0$, yields:

$$L_0/L_F = \tfrac{1}{2} + \tfrac{1}{2}[1 - \gamma_0]^{1/2}. \qquad \text{(Classical limit)} \qquad 24$$

This result is plotted in Fig. 4. $L_0/L_F$ displays a phase-fluctuation driven phase transition whose features are similar to the KTB transition. With the prefactor $a = ¼$, the transition would occur at $\gamma_0(T_{TPF}) = 1$ if it were not preceded by the KTB transition at $\gamma_0(T_{KTB}) \approx 0.90$. At the transition, $L_0/L_F = ½$ and $d(L_0/L_F)/dT = -\infty$, meaning that $L_0/L_F$ drops discontinuously from ½ to zero. The value (0.50) of $L_0/L_F$ at the transition is independent of $a$, and it is close to the values of $L_0/L_A$ (0.64, 0.60, and 0.54) at $T_{KTB}$ for honeycomb, square, and triangular arrays, respectively. Thus, Eq. (24) is physically reasonable.

On the basis of this analysis, we conclude that the effect of thermal phase fluctuations on the sheet inductance of films should be similar to their effect on arrays of Josephson junctions. Their effect should be small below $T_Q$ and increase rapidly as the 2-D transition is approached. Fluctuations should suppress $L_0/L_F$ by 20% to 30% just before the rapid drop which signals the 2-D transition. These conclusions are consistent with measurements on a-MoGe films.[5]

5. Disordered s-wave superconducting films.

When discussing films, it is common to discuss the 2-D penetration depth, $1/\lambda_\perp(T) \equiv d/\lambda^2(T) = \mu_0/L_F(T)$, rather than $L_F$. Dirty-limit s-wave superconductors are particularly simple. In them, the quantum crossover occurs at $\lambda_\perp(0)/\lambda_\perp(T_Q) \approx 1/7$ [Eq. (20)] because $R_N/L_F(0) \approx \pi\Delta(0)/\hbar$,[30] and $\Delta(0) \approx 2k_BT_{C0}$, leading to $T_0 = \pi\Delta(0)/k_B \approx 2\pi T_C$. The corresponding value of $T_Q$ is about 0.94 $T_{C0}$. As a practical matter, films which exhibit fluctuation effects large enough to study have sheet resistances, $R_N$, near 1 k$\Omega$, so $T_Q$ nearly coincides with $T_{KBT}$. Fluctuations turn on very rapidly with increasing T because quantum suppression diminishes as nonlinear effects turn on. As mentioned above, measurements on a-MoGe films are consistent with the model.[5] It remains to be seen whether other materials are consistent.

6. Clean d-wave superconductors: optimally doped YBCO.

Cuprates offer the opportunity to study thermal phase fluctuations in a clean quasi-2-D superconductor. Insofar as the G-L order parameter in cuprates is a complex scalar function, the foregoing analysis is applicable. Because of their sensitivity to disorder, d-wave superconductors require an extremely small elastic scattering rate, $\hbar/\tau_{el} \ll \Delta_0(0)/30$, to qualify as "clean" when strongly scattering impurities are present. The constraint lessens for weaker scatterers. For strongly-scattering impurities, the characteristic temperature, $k_BT^* \equiv [\hbar\Delta_0(0)/\tau_{el}]^{1/2}$ separates "very-low" temperatures from "low" temperatures.[31] The hallmark of clean cuprates is $\lambda^{-2}(T) - \lambda^{-2}(0) \propto T$ below about 0.3 $T_C$. Below $T^*$ impurity scattering causes a crossover from T-linear to $T^2$.

We are particularly interested in identifying the quantum crossover and examining behavior below that point. From Eq. (24), classical phase fluctuations lead to:

$$\lambda_{\perp,0}(T)/\lambda_\perp(T) \approx 1 - T/1500 \text{ K}, \qquad\qquad 25$$

with numbers appropriate for the *ab*-plane of optimally doped YBa$_2$Cu$_3$O$_{7-\delta}$: $\lambda_{ab}(0)$ = 150 nm [32] and $d$ = 1.17 nm,[33] so $\lambda_\perp(0)$ = 17 μm. $\lambda_{\perp,0}(T)$ is the mean-field penetration depth. Since $R_N \approx$ 100 μΩcm/1.17 nm ≈ 850 Ω, $T_0$ is about 300 K. From Eqs. (20) and (21), the quantum crossover occurs at $\lambda_\perp(0)/\lambda_\perp(T_Q) \approx$ ¼. For optimally-doped YBCO, this condition occurs at $T_Q \approx$ 0.90 - 0.95 $T_C$. Below $T_Q$, $\langle p_S^2 \rangle$ suppressed by a small factor:

$$k_B T\, G_F L_F/\hbar \approx [k_B T/\Delta_0(0)]^2. \qquad 26$$

But $a(T) \approx \Delta_0(T)/k_B T > 1$ in this regime, [34] reflecting the sensitivity of d-wave superconductors to superfluid motion, so γ is suppressed below its classical value by a single power of $k_B T/\Delta_0(T)$. With $\Delta_0(0)/k_B \approx$ 300 K, we find:

$$\lambda_{\perp,0}(T)/\lambda_\perp(T) \approx 1 - (T / 670\text{ K})^2. \qquad (\text{YBCO, } T < 40\text{ K.}) \qquad 27$$

Thus, thermal phase fluctuations cannot account for the linear T dependence of $1/\lambda_\perp(T)$ below 30 K. The observed linear behavior [11], $\lambda^2(0) d\lambda^{-2}(T)/dT \approx$ -1/180 K in optimally-doped YBCO is better interpreted as $-2\ell n 2 k_B/\Delta_0(0)$, yielding: $\Delta_0(0)/k_B \approx$ 250 K ≈ 3 $T_C$.

A more detailed analysis of cuprates, including optimally doped YBCO, slightly underdoped YBCO and optimally doped BSCCO will be presented elsewhere. [35]

7. Conclusion.

Guided by rigorous calculations of the inductances of resistively shunted Josephson junctions and 2-D arrays of junctions, the former presented as part of this work and the latter obtained from the literature, we have calculated the influence of thermal phase fluctuations on the superfluid density, or, magnetic penetration depth, of effectively 2-D superconductors. We find that thermal phase fluctuations are much more important than phase-slip fluctuations, except at temperatures very close to the super-to-normal transition. Quantum mechanics strongly suppresses phase fluctuations below a crossover temperature which is determined by the "R/L" low-pass frequency of the film, and which is expected to be above 0.9 $T_C$. There is experimental evidence for this crossover in measurements of the complex impedance of thin amorphous MoGe films.

Given that the quantum crossover is expected near $T_C$, thermal phase fluctuations cannot be responsible for the T-linear decrease in $\lambda^{-2}(T)$ at low T in optimally doped YBCO and BSCCO. At temperatures near $T_C$, the importance of thermal phase fluctuations in cuprates depends critically on the strength of interlayer coupling. More experimental and theoretical work is needed to pin down the systematics of phase fluctuations in conventional and cuprate superconductors.

*Acknowledgements*. This work was supported in part by DoE grant DE-FG02-90ER45427 through the Midwest Superconductivity Consortium. We appreciate helpful conversations with K.K. Likharev regarding calculations for a Josephson junction. We are grateful to Brent Boyce, David Stroud, and C. Jayaprakash for numerous discussions.

Figure Captions.

1. Circuit diagram for a model Josephson junction.

2. The quantum suppression factor, $f_Q \equiv \langle I_S^2 \rangle / \langle I_S^2 \rangle_C$, calculated from Eq. (8) for JJ's with $\omega_J \gg \omega_0$, $\omega_J = 0.7\,\omega_0$, and $\omega_J = \omega_0/4$. The dotted curves are approximations, Eqs. (9).

3. Normalized inductance, $L_J/L_0$, and resistance, $R_J/R$, *vs.* $\gamma_0(T)$ for an overdamped ($\omega_J \gg \omega_0$), resistively-shunted Josephson junction. $\omega_J \equiv (L_0 C)^{-1/2}$ and $\omega_0 \equiv R/L_0$. The uppermost curve for $L_J/L_0$ is for low frequency, $\omega/\omega_0 \ll 1$, while the uppermost curve for $R_J/R$ is for high frequency, $\omega/\omega_0 \gg 1$.

4. Normalized inverse inductances *vs.* $\gamma_0$, calculated in the classical limit for an overdamped, resistively-shunted JJ (solid curve), a square array of identical JJ's (connected dots), and a 2-D superconducting film (solid curve). The junction and array are well approximated at low $\gamma_0$ by: $L_0/L_J \approx (1 - \gamma_0)^{1/2}$ and $L_0/L_A \approx (1 - \gamma_0/2)^{1/2}$, (dotted curves), respectively.

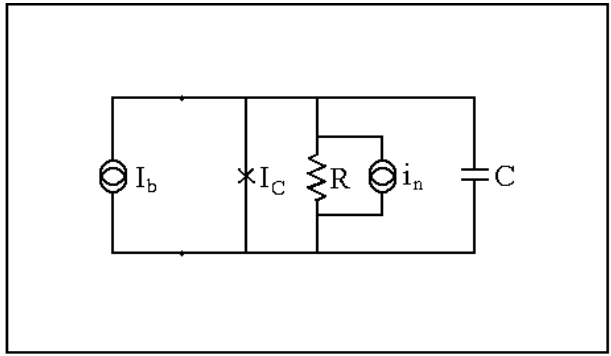

Figure 1

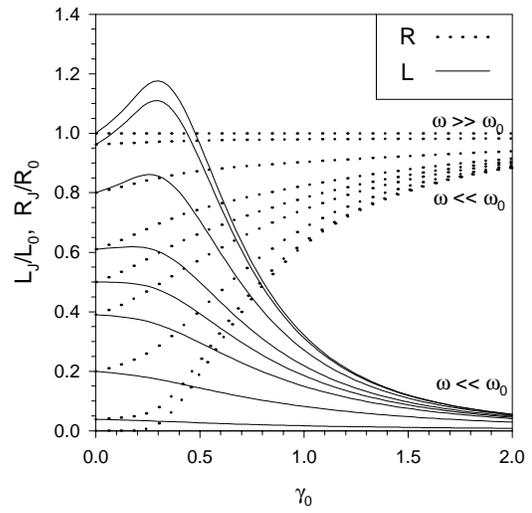

Figure 3

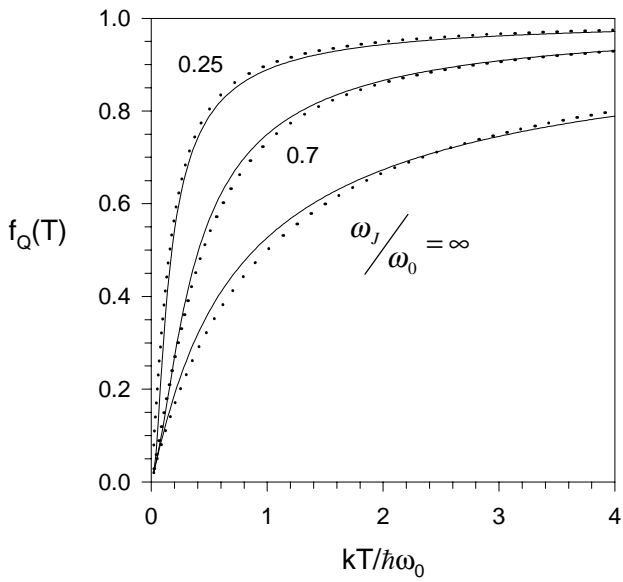

Figure 2

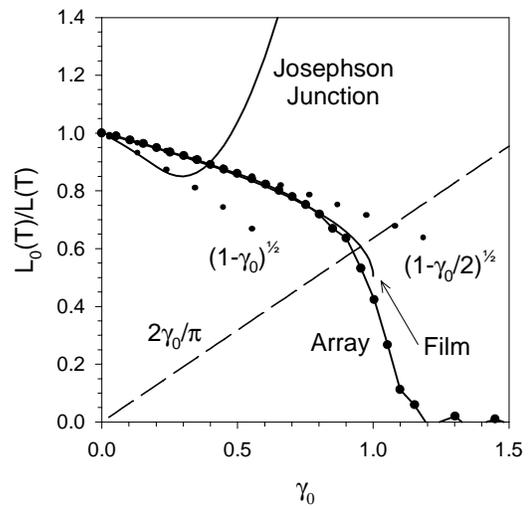

Figure 4